\documentclass[10pt,preprint]{emulateapj}

\slugcomment{Accepted to ApJ on 12 August 2010}

\shorttitle{$z=7$ Ly$\alpha$ Emitters in SXDS}
\shortauthors{Ota et al.}

\begin{document}


\title{Ly$\alpha$ Emitters at $z=7$ in the Subaru/{\it XMM-Newton} Deep Survey Field:\\
Photometric Candidates and Luminosity Functions\altaffilmark{1}}


\author{Kazuaki Ota\altaffilmark{2}, Masanori Iye\altaffilmark{3,4,5}, Nobunari Kashikawa\altaffilmark{3,4}, Kazuhiro Shimasaku\altaffilmark{5}, Masami Ouchi\altaffilmark{6,7}, Tomonori Totani\altaffilmark{8}, Masakazu A. R. Kobayashi\altaffilmark{3,9}, Masahiro Nagashima\altaffilmark{10}, Atsushi Harayama\altaffilmark{11}, Natsuki Kodaka\altaffilmark{11}, Tomoki Morokuma\altaffilmark{3,9}, Hisanori Furusawa\altaffilmark{3}, Akito Tajitsu\altaffilmark{12}, Takashi Hattori\altaffilmark{12}}

\email{ota@icrr.u-tokyo.ac.jp}


\altaffiltext{1}{Based on data collected at Subaru Telescope, which is operated by the National Astronomical Observatory of Japan.}

\altaffiltext{2}{Institute for Cosmic Ray Research, University of Tokyo, 5-1-5 Kashiwa-no-Ha, Kashiwa City, Chiba, 77-8582, Japan}

\altaffiltext{3}{National Astronomical Observatory of Japan, 2-21-1 Osawa, Mitaka, Tokyo, 181-8588, Japan}

\altaffiltext{4}{The Graduate University for Advanced Studies, 2-21-1 Osawa, Mitaka, Tokyo, 181-8588, Japan}

\altaffiltext{5}{Department of Astronomy, Graduate School of Science, University of Tokyo, 7-3-1 Hongo, Bunkyo-ku, Tokyo 113-0033, Japan}

\altaffiltext{6}{Observatories of the Carnegie Institution of Washington, 813 Santa Barbara Street, Pasadena, CA 91101, USA}

\altaffiltext{7}{Carnegie Fellow}

\altaffiltext{8}{Department of Astronomy, School of Science, Kyoto University, Sakyo-ku, Kyoto 606-8502}

\altaffiltext{9}{Research Fellow of Japan Society of Promotion of Science}

\altaffiltext{10}{Faculty of Education, Nagasaki University, Nagasaki, 852-8521, JAPAN}

\altaffiltext{11}{Department of Physics, Saitama University, Shimo-Okubo 255, Sakura, Saitama 338-8570, Japan}

\altaffiltext{12}{Subaru Telescope, National Astronomical Observatory of Japan, 650 North Afohoku Place, Hilo, HI 96720, USA}


\begin{abstract}
We conducted a deep narrowband NB973 (FWHM $=$ 200\AA~centered at 9755\AA) survey of $z=7$ Ly$\alpha$ emitters (LAEs) in the Subaru/{\it XMM-Newton} Deep Survey Field, using the fully depleted CCDs newly installed on the Subaru Telescope Suprime-Cam, which is twice more sensitive to $z=7$ Ly$\alpha$ at $\sim1 \mu$m than the previous CCDs. Reaching the depth 0.5 magnitude deeper than our previous survey in the Subaru Deep Field that led to the discovery of a $z=6.96$ LAE, we detected three probable $z=7$ LAE candidates. Even if all the candidates are real, the Ly$\alpha$ luminosity function (LF) at $z=7$ shows a significant deficit from the LF at $z=5.7$ determined by previous surveys. The LAE number and Ly$\alpha$ luminosity densities at $z=7$ is $\sim 7.7$--54\% and $\sim5.5$--39\% of those at $z=5.7$ to the Ly$\alpha$ line luminosity limit of $L({\rm Ly}\alpha)\gtrsim 9.2 \times 10^{42}$ erg s$^{-1}$. This could be due to evolution of the LAE population at these epochs as a recent galaxy evolution model predicts that the LAE modestly evolves from $z=5.7$ to 7. However, even after correcting for this effect of galaxy evolution on the decrease in LAE number density, the $z=7$ Ly$\alpha$ LF still shows a deficit from $z=5.7$ LF. This might reflect the attenuation of Ly$\alpha$ emission by neutral hydrogen remaining at the epoch of reionization and suggests that reionization of the universe might not be complete yet at $z=7$. If we attribute the density deficit to reionization, the intergalactic medium (IGM) transmission for Ly$\alpha$ photons at $z=7$ would be $0.4 \leq T_{{\rm Ly}\alpha}^{\rm IGM} \leq 1$, supporting the possible higher neutral fraction at the earlier epochs at $z>6$ suggested by the previous surveys of $z=5.7$--7 LAEs, $z\sim6$ quasars and $z>6$ gamma-ray bursts.
\end{abstract}





\section{Introduction} 
The recent progress on the observations of high redshift galaxies are really remarkable and has been revealing how galaxies evolved and contributed to reionization. Optical narrowband surveys have discovered Ly$\alpha$ emitters (LAEs) up to $z\simeq6$--7 \citep[e.g.,][]{Shima06,Tani05,kashik06,06IOK,Ouchi08,Ota08}. A few pioneering attempts to search for $z\sim9$ LAEs with near-infrared narrowband were made and ended with null detections \citep{Willis05,Cuby07}. Hight redshift LAEs can be a probe of reionization since observed Ly$\alpha$ line luminosity function (Ly$\alpha$ LF) is expected to decline beyond $z\sim6$, where reionization is thought to have completed based on the quasar and $\gamma$-ray burst (GRB) observations \citep{Fan06,Totani06,Greiner09}, as the increasing fraction of IGM neutral hydrogen absorbs or scatters the Ly$\alpha$ photons from LAEs \citep{HiSpa99,RM01,Hu02}. Earlier works showed that Ly$\alpha$ LF does not change much beyond $z\sim6$ by comapring LFs at $z=5.7$ and 6.5 and concluded that the universe was largely ionized at $z=6.5$ \citep{Stern05,MR04}. Also, some cosmological simulations of reionization support this result \citep[{\it e.g.},][]{HaimanCen05}. These studies used the samples compiled from various literatures with different selection criteria, survey areas and depths and suggested that the larger deeper uniform samples are necessary to improve the accuracy and statistics and draw the firmer conclusion. Later, \citet{Shima06} and \citet{kashik06} constructed a large homogeneous samples of $z=5.7$ and 6.6 selected from the consistent criteria, the same and large survey area and comparable depth to find the significant decline of Ly$\alpha$ LF from $z=5.7$ to 6.6 and suggested possible increase in neutral fraction at $z>6$ and the end of reionization at $z\simeq6$. For even earlier epoch, \citet{Hibon10} recently performed a near-infrared narrowband survey and detected seven $z=7.7$ LAE candidates. They concluded that Ly$\alpha$ LF does not evolve from $z=6.6$ to 7.7 if two brightest candidates in their sample are real but otherwise it does evolve, and emphasized that sample contamination, if any, easily affects the results and thus spectroscopic comfirmation is important. 

Meanwhile, broadband dropout techniques have detected Lyman break galaxies (LBGs) up to $z\sim 6$--10 \citep[e.g.,][]{Stanway07,Bouw05,Bouw06,Bouw09,Bouw10,Henry07,Henry08,Bunker09,Hickey09,Oesch09,Oesch10,Ouchi09b}. Their UV continuum LF (UV LF) implies rapid buildup of luminous galaxies from $z>7$ to $z\lesssim6$ \citep[e.g.,][]{06BI}. However, the faint-end slope of $z \gtrsim 6$ LBG UV LF and the escape fraction of ionizing photons are still highly uncertain, and galaxy contribution to reionization has not been tightly constrained yet \citep{Bunker04,YW04,Bouw06,Bouw08,Henry09}. 

Recently, gravitationally lensing galaxy clusters have been used to search for magnified extremely faint galaxies at $z>7$ and to probe faint end slopes of Ly$\alpha$ and UV LFs. Three types of ambitious challenges have been made lately. Using the Keck Telescope, \citet{Stark07} conducted near-infrared blind spectroscopic survey of $z\sim 8.5$--10.4 LAEs by placing long slits to the critical lines, locations of the highest magnification, of 9 lensing clusters, and found six candidates. On the other hand, $z>7$ LBGs were surveyed over several lensing clusters using deep optical--infrared imagings of {\it Hubble Space Telescope} (HST), {\it Spitzer} and VLT, covering most of the magnification regions including critical lines \citep{Bouw09,Bradley08,Richard08}. \citet{Richard08} detected 10 $z\sim7$--8 $z$-dropouts and 2 $z\sim9$--10 $J$-dropouts. Both \citet{Stark07} and \citet{Richard08} found that the number densities of very faint LAEs and LBGs at $z>7$ are quite high, and they can supply the necessary photons to sustain reionization. However, their faintness has not yet enabled successful spectroscopic identification of them, and they still stay candidates. Finally, \citet{Willis08} conducted a narrowband survey of $z\sim9$ LAEs targeting 3 lensing clusters and ended with no detection of candidates.  

Most of the insights to the galaxy evolution and reionization at $z \gtrsim 7$ are based on candidate galaxies. We do not know how many of them are real, and this fraction easily affects the analysis results. Hence, it is important to construct samples of low contamination and of spectroscopically confirmed galaxies at $z \gtrsim 7$. To this end, we conducted a narrowband NB973 ($\lambda_c=9755$\AA~and $\Delta\lambda=200$\AA~corresponding to $6.94 \leq z \leq 7.11$ Ly$\alpha$ emission) imaging of the Subsru Deep Field \citep[SDF, 876arcmin$^2$;][]{kashik04} with Suprime-Cam \citep{Miyazaki02} on the Subaru Telescope and successfully identified a $z=6.96$ LAE by spectroscopy \citep{06IOK}. The $z\simeq7$ LAE number density was only 17\% of that at $z=6.6$ with an estimate of lower intergalactic medium (IGM) transmission for Ly$\alpha$ photons at higher redshifts; $T_{{\rm Ly}\alpha}^{\rm IGM} \sim 0.62$--0.78 at $z=6.6$ and $T_{{\rm Ly}\alpha}^{\rm IGM} \sim 0.40$--0.64 at $z=7$ \citep{Ota08}. This series of density declines and decrease in $T_{{\rm Ly}\alpha}^{\rm IGM}$ at $z\gtrsim6$ could be another evidence that reionization ended at $z\sim6$. 

\begin{figure}
\plotone{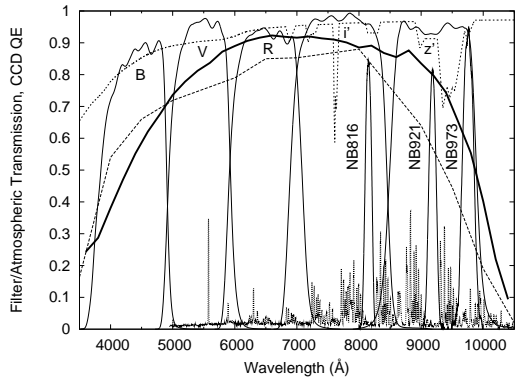}
\caption{Filter transmission of the Suprime-Cam broadbands $BVRi'z'$ and narrowbands NB816, NB921 and NB973 (thin solid curves) used for our photometry. The OH night skylines are also overplotted with thin dotted curve. The short-dashed curve is the atmospheric transmission. The long-dashed and thick solid curves show the quantum efficiency of the old CCDs and the new fully depleted CCDs. The new CCDs are about twice more sensitive to the $z=7$ Ly$\alpha$ at $\sim 9730$\AA. \label{MS_BVRizNBfilters_z7LAE}}
\end{figure}

However, our conclusion from the $z=7$ LAE was naive because of three weaknesses. First of all, the survey limit was as shallow as the Ly$\alpha$ luminosity of $L({\rm Ly\alpha}) \gtrsim 10^{43}$ erg s$^{-1}$, mostly due to the low sensitivity of detector at the wavelength $\sim 1\mu$m. Hence, our study was limited to the very bright end of LF. Secondly, we could detect only one $z=7$ LAE and thus the sample was statistically very small. Of course, this might be really due to the significant amount of absorption of Ly$\alpha$ photons by neutral hydrogen that might remain at $z=7$, but we cannot completely deny the possibility of the small number detection due to the shallow survey depth and the low detector sensitivity. Finally, we observed only one sky field, the SDF. If we observe any different sky fields, we might obtain different results due to the possible cosmic variance of galaxy evolution and reionization state. To discuss galaxy evolution and reionization more robustly, we need a larger $z=7$ LAE sample from a deeper survey than the SDF over another comparably large sky area.

The chance to overcome these weaknesses had come in 2008. The CCDs of Suprime-Cam were replaced with new CCDs that adopt a much larger depletion layer, allowing them to detect more photons at redder wavelength. The CCDs have twice better sensitivity to the $z=7$ Ly$\alpha$ emission at $\sim1\mu$m than the previous CCDs. Inspired with this, we have performed a new deeper NB973 filter survey of $z=7$ LAEs, targeting a different but comparably large area sky field, the Subaru/{\it XMM-Newton} Deep Survey Field \citep[SXDS;][]{Furusawa08}. We have detected more candidate $z=7$ LAEs to 0.5 mag deeper limit than our previous survey in the SDF. In the current paper, we explore the photometric properties of them and obtain any implication for reionization based on the deeper Ly$\alpha$ LF considering the cosmic variance. The paper is organized as follows. In \textsection 2, we describe the properties of the new CCDs and imaging observation of the SXDS. Then, the $z=7$ LAE candidates are selected and their photometric properties are analyzed in \textsection 3. In \textsection 4, we derive the deeper Ly$\alpha$ LF based on the newly detected LAE candidates, compare it with LFs from other recent studies and discuss its implication for galaxy evolution and cosmic reionization. We conclude and summarize our results in \textsection 5. Throughout, we adapt a concordance cosmology with $(\Omega_m, \Omega_{\Lambda}, h)=(0.3, 0.7, 0.7)$, and AB magnitudes with $2''$ diameter aperture, unless otherwise specified.       

\section{Observation}
The new fully depleted red-sensitive CCDs were installed in July 2008 on the Suprime-Cam. The quantum efficiency of the previous and new CCDs as well as transmission curve of the NB973 filter are shown in Figure 1. The quantum efficiency of the new CCDs at the central wavelength of NB973, 9755\AA, is $\sim0.6$, about twice better than that of the previous CCDs. This substantially reduces the integration time required to reach a deeper limit than the SDF NB973 survey. 

We targeted another sky region studied by the Subaru Observatory Project, the SXDS field \citep{Furusawa08}, because deep optical broadband $BVRi'z'$ images taken by the Suprime-Cam are available. Also, deep narrowband NB816 ($\lambda_c=8160$\AA~and $\Delta\lambda_{\rm FWHM}=$120\AA~corresponding to $z\sim5.7$ Ly$\alpha$) and NB921 ($\lambda_c=9196$\AA~and $\Delta\lambda_{\rm FWHM}=$132\AA~corresponding to $z\sim6.6$ Ly$\alpha$) filter images of the SXDS were taken by \citet{Ouchi08,Ouchi09a}. The entire SXDS field has an area of $\sim 1.3$ square degree, consisting of five pointing of Suprime-Cam. They are named as SXDS-C, SXDS-N, SXDS-S, SXDS-E and SXDS-W with the central coordinates 02$^{\rm h}$18$^{\rm m}$00.$^{\rm s}$0, -05$^o00'00''$ (J2000) of the SXDS-C corresponding to the center of the entire SXDS field. Moreover, this field has been observed with several different wavelength from X-ray to radio. Hence, multi-wavelength study of $z=7$ LAEs is also possible. For example, the SXDS-C and most part of the SXDS-S have been observed with the UKIDSS Ultra Deep Survey \citep[UKIDSS-UDS;][]{Lawrence07} and the Spitzer legacy survey (SpUDS; PI: J. Dunlop)\footnote[13]{\tiny http://ssc.spitzer.caltech.edu/spitzermission/observingprograms/legacy/}. However, the SXDS-C includes a few bright stars whose stellar halos contaminate large areas. Hence, to image the same area as the UKIDSS and the Spitzer surveys and to avoid those bright stars, we observed an area between the SXDS-C and SXDS-S, 02$^{\rm h}$18$^{\rm m}$00.$^{\rm s}$0, -05$^o13'30''$ (J2000), as shown in Figure 2 by one pointing ($34' \times 27'$) of the Suprime-Cam. 

The NB973 imaging observations with the new red-sensitive CCDs were carried out at dark clear nights on 25 and 26 October and 30 November in 2008. The sky conditions were photometric with the seeing of $\sim0.''5$--$1.''2$. By taking the 30 minute exposure frames with 8-point dithering patterns, we obtained a total of 17 hours of imaging data. We have reduced the NB973 image frames using the software SDFRED \citep{Ouch04,Yagi02} in the same standard manner as in \citet{kashik04} and \citet{Ota08}. The dithered NB973 image frames were combined by discarding the frames with seeing size of $>1''$ because we confirmed that they degrade the final stacked image although they do not help improve the depth. The astrometry of the combined image was then matched to those of the $BVRi'z'$ SXDS-C and SXDS-S images to produce the two images we hereafter call NB973-SXDS-C and NB973-SXDS-S. The seeing size and integration time of these final combined NB973 images are $1.''0$ and 13 hours. Spectrophotometric standard stars LDS749B, GD248 and GD71 \citep{Oke90} were imaged during the observation to calibrate the photometric zeropoint of these stacked images, which was turned out to be NB973 $=32.24$ mag ADU$^{-1}$. The limiting magnitude reached NB973 $\leq25.4$ at $5\sigma$ with 13 hour integration, about 0.5 magnitude deeper than our previous 15 hours of the SDF NB973 survey. 

\begin{figure}
\plotone{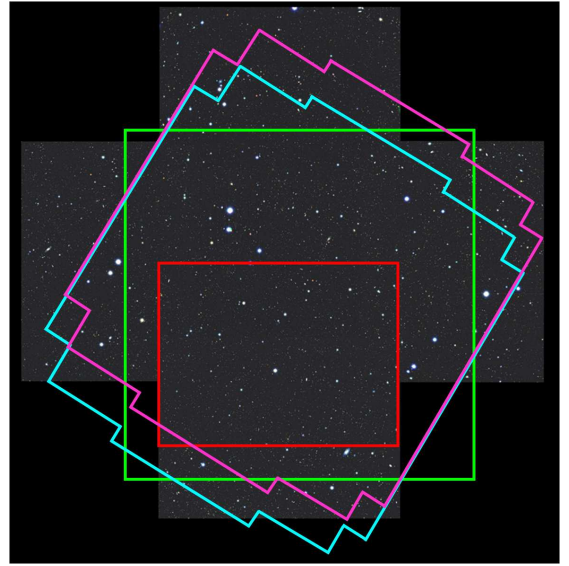}
\caption{The image of the Subaru/{\it XMM-Newton} Deep Survey (SXDS) Field \citep{Furusawa08}. The north is up and the east to the left. The SXDS consists of five pointing (totally $\sim 1.3$ square degree) of the Suprime-Cam named SXDS-C, SXDS-N, SXDS-S, SXDS-E and SXDS-W. We imaged the region between the SXDS-C and SXDS-S with the NB973 (the red square). The green, cyan and magenta lines show the UKIDSS-UDS \citep{Lawrence07} and SpUDS fields (PI: J. Dunlop) with {\it Spizer Space Telescope} Infrared Array Camera $3.6\mu$m and $5.8\mu$m imaging and $4.5\mu$m and $8.0\mu$m imaging, respectively. \label{SXDS-C-S-UKIDSS-Spitzer}}
\end{figure}



\section{Photometric Candidates}

\subsection{Photometry}
Using the NB973-SXDS-C and NB973-SXDS-S images, we made an NB973-detected object catalog and performed photometric analysis. Because the broadband $BVRi'z'$ and narrowband NB816 and NB921 filter images of the SXDS-C and SXDS-S have smaller seeing sizes of $\sim 0.''7$--$0.''8$, we convolved these images to have the same seeing size of $1''$ as the NB973-SXDS-C and NB973-SXDS-S for the purpose of aperture photometry. The $2''$ aperture limiting magnitudes of these images at $3\sigma$ with the Galactic extinction corrected based on \citet{Schlegel98} are $(B, V, R, i', z', {\rm NB816, NB921})=(28.00, 27.71, 27.52, 27.58, 26.54, 26.52, 26.37)$ for the SXDS-C and $(28.24, 27.68, 27.62, 27.43, 26.36, 26.62, 26.37)$ for the SXDS-S. We used the SExtractor software version 2.2.2 \citep{BA96} for source detection and photometry. The Suprime-Cam CCDs have a pixel size of $\sim 0.''2$ pixel$^{-1}$. We considered an area larger than contiguous 5 pixels with a flux [mag arcsec$^{-2}$] greater than $2\sigma$ to be an object. Object detection was first made in the NB973-SXDS-C and NB973-SXDS-S images and then photometry was done in the images of the other wavebands for SXDS-C and SXDS-S, respectively, using the double-image mode. We measured $2''$ diameter aperture magnitudes of detected objects with MAG$_{-}$APER parameter and total magnitudes with MAG$_{-}$AUTO. The final object catalog was constructed by combining the photometry from both SXDS-C and SXDS-S, detecting the total of 230,044 objects down to NB973 $\leq25.4$ ($5\sigma$).    

\begin{figure}
\plotone{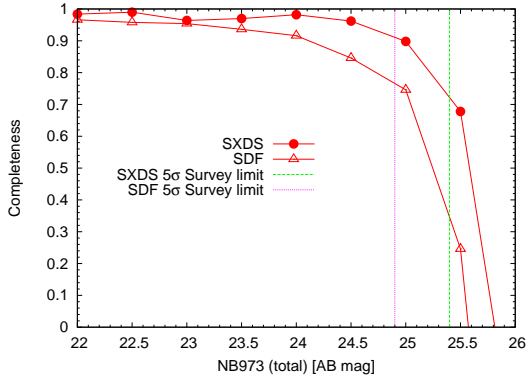}
\caption{The detection completenesses of the NB973 SXDS (the present survey) and SDF \citep[the previous survey][]{Ota08} images, calculated for every 0.5 mag bin. The completenesses do not reach 1.0 even for the objects with bright magnitudes. This is because the blended or overlapped objects tend to be counted as one object by the SExtractor. The completenesses are corrected when the number and luminosity densities of $z=7$ LAEs in the SXDS and SDF are calculated in \textsection 4. Our current survey in the SXDS is much deeper and more complete than our previous survey in the SDF. \label{MS_AVE_NB973_Completeness}}
\end{figure}

\subsection{Selection of Candidate $z=7$ LAEs \label{COLOR_CRITERIA}}
We used the same color criteria as ones adopted for our previous NB973 survey in the SDF \citep{Ota08} to select out candidate $z=7$ LAEs in the SXDS. Namely, 
\begin{itemize}
\item [(1)] $B, V, R, i', {\rm NB816}, z', {\rm NB921} < 3\sigma$; ${\rm NB973} \leq 25.4$
\item [(2)] $B, V, R, {\rm NB816}, {\rm NB921} < 3\sigma$; $i'-z'>1.3$;\\
$z'- {\rm NB973} > 1.0$; ${\rm NB973} \leq 25.4$
\end{itemize}   
We found 1274 and 18 objects satisfying the criterion (1) and (2), respectively. 

However, most of them are located in the noisy reagions near the edges of the NB973-SXDS-C and NB973-SXDS-S images where the signal-to-noise (S/N) ratio is low. This implies that most of them could be noises that appear only in the NB973 images and not in the other wavebands. To examine if they are noises, we created the negative NB973-SXDS-C and NB973-SXDS-S images by multiplying each pixel value by $-1$ and performed source detection using SExtractor with the detection and analysis thresholds of $5\sigma$, instead of $2\sigma$, because our candidate selection adopted ${\rm NB973} \leq 25.4$ ($5\sigma$) and the sky background rms's do not change even if we change the positive images into negative ones. The top and right edges of the NB973-SXDS-C image and the bottom and right edges of the NB973-SXDS-S images were dominated with negative $> 5\sigma$ detections, which are considered noises. These edge regions coincide with the locations where most of the sources selected with the $z=7$ LAE candidate criteria (1) and (2) above distribute. Hence, we trimmed these edge regions off the NB973-SXDS-C and -S images. After that, we also measured the limiting magnitudes near the edge of the images and found that the regions within 1000 pixels ($\sim 3.'4$) from right edges of the NB973-SXDS-C and -S images are 0.3 mag shallower. Thus, we also removed these region from the images. 


\begin{figure}
\plotone{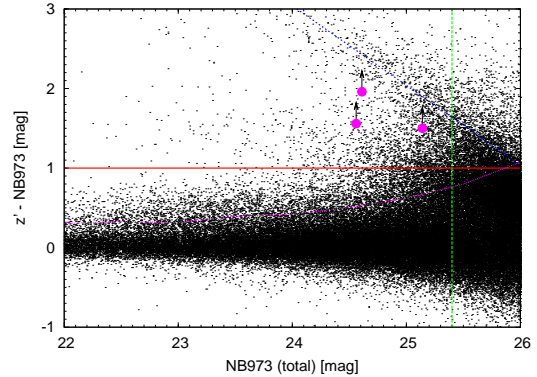}
\caption{$z'-$ NB973 ($2''$ aperture mags.) color as a function of NB973 (total) magnitude of the objects detected in our survey field (shown by dots). The dotted curve shows $3\sigma$ error track of $z'-$ NB973 color.  The horizontal solid line is a part of our color selection criterion, $z'-$ NB973 $>1.0$. The vertical dashed line indicates the detection limiting magnitude of our survey, NB973 $=25.4 (5\sigma)$. The diagonal dashed line is the $2\sigma$ limits of $z'$-band $2''$ aperture magnitude. The three $z=7$ LAE candidates are denoted by the filled circles with the upper arrows showing the $2\sigma$ limits on $z'-$ NB973 colors.\label{MS_CMD}}
\end{figure}

Consequently, we were left with three objects located in NB973-SXDS-S image, all of which satisfy the candidate selection criterion (1). We visually inspected all the broadband and narrowband images of each of these three objects to remove obviously spurious sources such as columns of bad pixels, pixels saturated with bright stars, noise events of deformed shapes and scattering pixels having anomalously large fluxes. None of the these applied to the three objects. We also examined if the objects are not seen in the $BVRi'$, NB816 and NB921 because such objects are considered low-$z$ interlopers since flux at the wavelength shortward $z=7$ Ly$\alpha$ should be mostly absorbed by the IGM. None of the three were seen in any of the wavebands blueward $z=7$ Ly$\alpha$. 

\begin{figure*}
\plotone{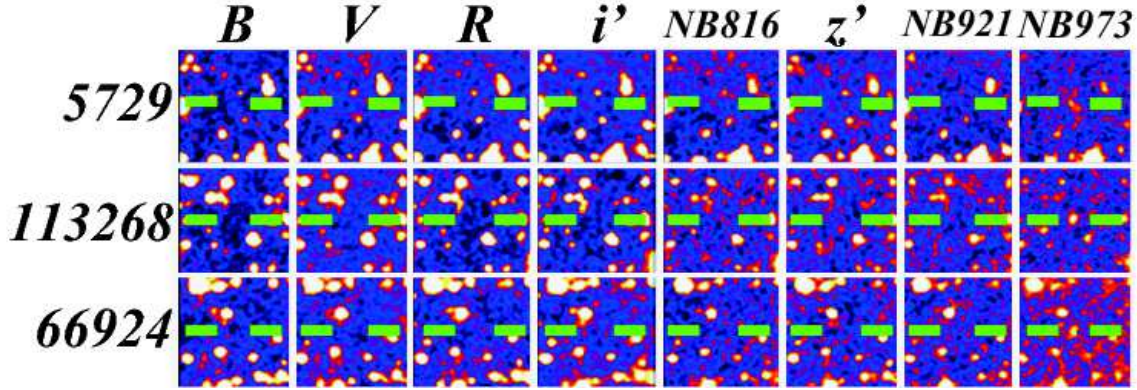}
\caption{The multi-waveband images of the three candidate $z=7$ LAEs. The size of each image is $20'' \times 20''$. They are detected only in NB973. \label{MS_Poststamp_IOK1-5}}
\end{figure*}

However, because these selected objects are only detected in the NB973 image and not in the other wavebands, they might include transient or variable objects such as active galactic nuclei (AGNs) and supernovae that were too faint to be detected at the time of $BVRi'z'$, NB816 and NB921 imaging observations of the SXDS but became brighter than $5\sigma$ at the time of our NB973 imaging. Also, the three objects might still include noise events that exist only in the NB973 image and that resemble real objects in shape. The color selection and the visual inspection above alone cannot remove these contaminants. 

To reduce the number of these contaminants as much as possible, we conducted another type of photometric analysis as follows. We created three additional NB973 SXDS images: (1) a stacking of NB973 image frames (11 hours in total and PSF of $1.''2$) taken on 25 and 26 October 2008, (2) a stacking of NB973 image frames (6 hours in total and PSF of $1.''2$) taken on 30 November 2008 and (3) a stacking of all the NB973 image frames including the ones with PSF $>1''$ (17 hours in total and PSF of $1.''2$) taken in both October and November 2008. Then, we measured the $2''$ aperture fluxes of the three objects in each of the images (1)--(3). We found that each object was detected in all the images (1)--(3) at 4.3--5.0$\sigma$, 5.0--7.8$\sigma$ and 3.0--4.4$\sigma$ significances, respectively. On the other hand, we also created the negative images from the images (1)--(3), performed source detection in the same way above and found that negative $>5\sigma$ noises were not detected in all the images (1)--(3) in the same locations.
Hence, the three objects are not likely to be noises. Meanwhile, the difference in NB973 flux between the images (1) and (2) ({\it i.e.}, between the October and November 2008) in each of the three objects was 0.32$\sigma$, 1.7$\sigma$ and 0.66$\sigma$, respectively. Since the flux differences are within $\pm 1\sigma$ photometric errors, all the three objects did not show any significant variability in one month. Hence, they might not be variable objects. However, we cannot completely exclude the possibility of them being variable objects because their time variability might be longer than one month. 

Eventually, we were left with three candidates and this is an upper limit on the number of $z=7$ LAEs in the NB973 SXDS field since they might still include variable objects with slow variability. In our previous NB973 survey in the SDF, \citet{Ota08} estimated that the number of variables could be $\sim3.8$--4.5 in the 876 arcmin$^2$ of sky area. If this number is correct and also applies to the NB973-SXDS-C and NB973-SXDS-S images (totally 824 arcmin$^2$, see below), all the three objects selected in the SXDS could be variable objects. However, to securely identify them, the follow-up spectroscopy is of course required. The color-magnitude diagram ($z'-$ NB973 vs. NB973) of the three LAE candidates and all the objects detected down to NB973 $=25.4$ is plotted in Figure 4. The images of the three LAE candidates and their photometric properties are shown in Figure 5 and Table \ref{Photo-property}, respectively.

To remove the noises in the process of the candidate selection, we trimmed the low S/N edge regions off the NB973-SXDS-C and NB973-SXDS-S images. After this, the total area of the NB973-SXDS-C and NB973-SXDS-S images became 824 arcmin$^2$. The comoving distance along the line-of-sight corresponding to the redshift range $6.94 \leq z \leq 7.11$ for LAEs covered by NB973 filter is 58 Mpc. Therefore, we have probed a total of $3.0 \times 10^5$ Mpc$^3$ volume in our survey.

\begin{table*}
\begin{center}
\caption{Photometric properties of candidate $z=7$ Ly$\alpha$ emitters\label{Photo-property}}
\begin{tabular}{lcccccccc}
\tableline\tableline
Object & R.A. (J2000.0)	& Decl. (J2000.0) & $i'$ & NB816 & $z'$ & NB921 & NB973 & NB973 (total) \\
\tableline
NB973-SXDS-S-5729	& 02:17:48.97 & -05:27:18.56 & $>27.87$ & $>27.06$ & $>26.80$ &	$>26.81$ &	25.24 &	24.52 \\
NB973-SXDS-S-113268	& 02:17:59.54 & -05:14:07.61 & $>27.87$ & $>27.06$ & $>26.80$ &	$>26.81$ &	24.84 &	24.57 \\
NB973-SXDS-S-66924	& 02:17:57.85 & -05:18:47.46 & $>27.87$ & $>27.06$ & $>26.80$ &	$>26.81$ &	25.30 &	25.10 \\
\tableline
\end{tabular}
\end{center}
\vspace*{-0.2cm}
NOTE: Units of coordinate are hours: minutes: seconds (right ascension) and degrees: arcminutes: arcseconds (declination) using J2000.0 equinox.  $i'$, NB816, $z'$, NB921 and NB973 are all $2''$ aperture magnitudes while NB973 (total) is a total magnitude.  Magnitudes are replaced by their $2\sigma$ limits if they are fainter than the limits.
\end{table*}

\subsection{The Detection Completeness}
The number of objects we can detect decreases as the fluxes of objects become fainter. To examine what fraction of objects in the NB973 image the SExtractor can detect or fails to detect down to the limiting magnitude of NB973 $\leq 25.4$, we measured the detection completeness of our photometry for the NB973-SXDS-C and NB973-SXDS-S images in the same way as \citet{Ota08}. The detection completeness is the ratio of the number of detected objects to that of objects actually imaged in the NB973-SXDS-C and NB973-SXDS-S images. The result is shown in Figure \ref{MS_AVE_NB973_Completeness}. The completeness is $\sim 72$\% at our detection limit of NB973 $=25.4$. This completeness was corrected when the number and luminosity densities and Ly$\alpha$ LF of $z=7$ LAEs were calculated in the \textsection 4. Figure \ref{MS_AVE_NB973_Completeness} also shows the completeness of our previous NB973 survey in the SDF for comparison. It is clear that the present survey in the SXDS is much deeper and more complete.

\subsection{Ly$\alpha$ Luminosity and Star Formation Rate\label{PHOT_LAE_Model}}
To derive the Ly$\alpha$ LF of $z=7$ LAEs, we need to know their Ly$\alpha$ fluxes. Now that we know the NB973 magnitudes of our $z=7$ LAE candidates, we can estimate their Ly$\alpha$ fluxes, $F({\rm Ly}\alpha)$, Ly$\alpha$ luminosities, $L({\rm Ly}\alpha)$, and corresponding star formation rates (SFRs), SFR$({\rm Ly}\alpha)$, by directly converting the NB973 magnitudes into $F({\rm Ly}\alpha)$. Note that these quantities could be upper limits because the narrowband NB973 filter might possibly detect not only the Ly$\alpha$ emission of a LAE but also its UV continuum flux if it is strong. In our present survery, we have neither sufficiently deep near-infrared images nor spectroscopic data for our three $z=7$ LAEs for the estimation of UV continuum fluxes to correct the narrowband fluxes for the continuum contribution. The $F({\rm Ly}\alpha)$, $L({\rm Ly}\alpha)$ and SFR$({\rm Ly}\alpha)$ calculated here are listed in Table \ref{Lumi-property}. We convert $F({\rm Ly}\alpha)$ into $L({\rm Ly}\alpha)$ by assuming that the redshift of Ly$\alpha$ emission is $z=7.02$ corresponding to the central wavelength of NB973 (9755\AA). The uncertainty of $L({\rm Ly}\alpha)$ due to the location of Ly$\alpha$ emission ($z=6.94$--7.10) within the NB973 passband ($\lambda=9655$--9855\AA) is $\Delta \log L({\rm Ly}\alpha)\lesssim 0.02$. To estimate SFR$({\rm Ly}\alpha)$, we used the following relation derived from Kennicutt's equation \citep{Kenicutt98} with the case B recombination theory \citep{Brockle71}.
\begin{equation}
{\rm SFR}({\rm Ly}\alpha) = 9.1 \times 10^{-43} L({\rm Ly}\alpha) M_{\odot} {\rm yr}^{-1}
\label{L-to-SFR_conversion}
\end{equation}
It should be noted that this relation holds good in the case where Ly$\alpha$ emission is not attenuated by neutral IGM. Ly$\alpha$ emission could be attenuated by neutral IGM at $z>6$. Hence, we basically discuss $L({\rm Ly}\alpha)$ and its volume density $\rho_{{\rm Ly}\alpha}$ as physical properties of $z=7$ LAEs in the subsequent sections and list SFR$({\rm Ly}\alpha)$ and its volume density SFRD$_{{\rm Ly}\alpha}$ just for reference.

\section{Discussion \label{Re-and-GalEv}}
In our previous survey, we detected one $z\simeq7$ LAE to the limit of $L({\rm Ly}\alpha) \gtrsim 1\times 10^{43}$ erg s$^{-1}$ in the SDF. The number and Ly$\alpha$ luminosity densities of $z=7$ LAE were significantly smaller than those of LAEs at $z=5.7$ and $z=6.6$, suggesting that neutral fraction might be larger at higher redshifts beyond $z\sim6$. However, the SDF survey had three weaknesses: shallow depth, small statistics and observation of only one sky field \citep{Ota08}. We have now detected three LAE candidates in a different sky field, SXDS, to 0.5 mag deeper limit than the SDF survey. Hence, we derive a deeper Ly$\alpha$ LF of $z=7$ LAEs with improved statistics to investigate what it implies for galaxy evolution and reionization and to see if it supports our previous result or how the result changes in the different sky field.
   
\subsection{Ly$\alpha$ Luminosity Function of $z=7$ LAEs \label{LyaLF}}
In \textsection \ref{PHOT_LAE_Model}, we estimated the upper limit on $L({\rm Ly}\alpha)$ for each LAE candidate. Based on this, we now derive the upper limit on Ly$\alpha$ LF of $z=7$ LAEs at $z=7$ to our survey limit of NB973 $\leq 25.4$ corresponding to $L({\rm Ly}\alpha) \gtrsim 9.2 \times 10^{42}$ erg s$^{-1}$ (also assuming that Ly$\alpha$ emission is at $z=7.02$ corresponding to the central wavelength of NB973 filter 9755\AA). We estimate the number density of LAEs by simply dividing the cumulative number of LAEs in each $L({\rm Ly}\alpha)$ bin by the survey volume of $3.0 \times 10^5$ Mpc$^3$ estimated in \textsection 2. For the error of number density in each bin, we include Poisson errors and cosmic variance estimated in the same way as in \citet{Ota08}. That is, we used the Tables 1 and 2 in \citet{Geh86} for the estimate of Poisson errors and the semi-analytic model of \citet{Some04} for the cosmic variance. Also, we correct the number density and its error for the detection completeness estimated in \textsection 2 and Figure 3. The derived upper limit Ly$\alpha$ LF of $z=7$ LAEs in the SXDS as well as LFs of LAEs at $z=5.7$, 6.6 and 7.0 in the SDF or SXDS from \citet{Shima06}, \citet{Ouchi08}, \citet{kashik06} and \citet{Ota08} are shown in Figure \ref{Fig_UpperLyaLF}. It should be noted that while our LF is the upper limit without correction for contribution from UV continuum flux, all the other LFs are corrected for this using either the deep broadband images redward Ly$\alpha$ or spectroscopy data (See each reference).

\begin{table*}
\begin{center}
\large
\caption{Luminosities and SFRs of candidate $z=7$ Ly$\alpha$ emitters\label{Lumi-property}}
\begin{tabular}{lccc}
\tableline\tableline
Object & $F({\rm Ly}\alpha)^a$ & $L({\rm Ly}\alpha)^b$ & SFR$({\rm Ly}\alpha)^c$ \\
 & ($10^{-17}$ erg s$^{-1}$ cm$^{-2}$) & ($10^{42}$ erg s$^{-1}$) & ($M_{\odot}$yr$^{-1}$) \\    
\tableline
NB973-SXDS-S-5729	& 3.6 &	$20.7\pm0.6$ & $18.8\pm0.5$ \\
NB973-SXDS-S-113268	& 3.4 &	$19.7\pm0.5$ & $18.0\pm0.5$ \\
NB973-SXDS-S-66924	& 2.1 &	$12.1\pm0.3$ & $11.0\pm0.3$ \\
\tableline
\end{tabular}
\end{center}
{$^a$The Ly$\alpha$ flux calculated assuming that $F({\rm Ly}\alpha) = F({\rm NB973})$.}\\
{$^b$The Ly$\alpha$ luminosity converted from $F({\rm Ly}\alpha)$ assuming that Ly$\alpha$ emission is located at the center of NB973 passband ({\it i.e.}, $z=7.02$). The uncertainties correspond to the other locations of Ly$\alpha$ emission ($z=6.94$--7.10) within the NB973 passband ($\lambda=9655$--9855\AA).}\\
{$^c$The SFR corresponding to $L({\rm Ly}\alpha)$.}
\end{table*}
 
The upper limit $z=7$ LF is similar to $z=5.7$ LFs at its brighter end but shows deficit from $z=5.7$ at the fainter luminosity. The fainter part of the upper limit $z=7$ LF is similar to the $z=6.6$ LF. The number density of the SXDS upper limit $z=7$ LF is larger than that of the SDF $z=7$ LF based on the spectroscopy. However, Figure \ref{Fig_UpperLyaLF} also shows that if we assume that $F({\rm NB973})=F({\rm Ly}\alpha)$ for the SDF $z=7$ LF, it is similar to the SXDS $z=7$ LF. This suggests that the field-to-field variance in $z=7$ Ly$\alpha$ LF might not be so significant between the SDF and SXDS. However, to investigate the field-to-field variance more precisely, we should know accurate Ly$\alpha$ luminosities of the three $z=7$ LAE candidates with spectroscopy. 

Figure \ref{Fig_MadauPlot} and Table \ref{Density} show the LAE number density $n_{{\rm Ly}\alpha}$ and Ly$\alpha$ luminosity density $\rho_{{\rm Ly}\alpha}$ estimated from our three $z=7$ LAE candidates to the survey limit of $L({\rm Ly}\alpha) = 9.2 \times 10^{42}$ erg s$^{-1}$. They are $n_{{\rm Ly}\alpha}^{z=7} \sim 1.9_{-0.9}^{+0.9} \times 10^{-5}$ Mpc$^{-3}$ and $\rho_{{\rm Ly}\alpha}^{z=7} \sim 1.9_{-0.9}^{+0.9} \times 10^{38}$ erg s$^{-1}$ Mpc$^{-3}$, respectively. Also, in Figure \ref{Fig_MadauPlot} and Table \ref{Density} to the same limit, the number and Ly$\alpha$ luminosity densities at $z=5.7$ and $z=6.6$ are estimated to be ($n_{{\rm Ly}\alpha}^{z=5.7{\rm phot}}, n_{{\rm Ly}\alpha}^{z=5.7{\rm phot}}, n_{{\rm Ly}\alpha}^{z=6.6{\rm phot}}, n_{{\rm Ly}\alpha}^{z=6.6{\rm spec}}) \sim (8.8_{-3.6}^{+4.2}, 6.8_{-1.5}^{+1.5}, 2.5_{-1.4}^{+1.6}, 1.1_{-0.8}^{+1.1}) \times 10^{-5}$ Mpc$^{-3}$ and $(\rho_{{\rm Ly}\alpha}^{z=5.7{\rm phot}}, \rho_{{\rm Ly}\alpha}^{z=5.7{\rm phot}}, \rho_{{\rm Ly}\alpha}^{z=6.6{\rm phot}}, \rho_{{\rm Ly}\alpha}^{z=6.6{\rm spec}}) \sim (12.4_{-5.1}^{+5.9}, 9.0_{-1.9}^{+1.9}, 2.9_{-1.6}^{+1.9}, 1.5_{-1.2}^{+1.5}) \times 10^{38}$ erg s$^{-1}$ Mpc$^{-3}$ from the $z=5.7$ Ly$\alpha$ LFs based on photometric LAE samples in SDF and SXDS derived by \citet{Shima06} and \citet{Ouchi08} and the $z=6.6$ Ly$\alpha$ LFs based on photometric and spectroscopic LAE samples derived by \citet{kashik06}, respectively. These densities were calculated by integrating the best-fit \citet{Schechter76} LFs derived in these studies and each uncertainty consists of Poisson error and cosmic variance. Again, we used the Tables 1 and 2 in \citet{Geh86} for the estimate of Poisson errors and the semi-analytic model of \citet{Some04} for the cosmic variance. Similarly, densities at $3<z<5.7$ were also calculated using the best-fit Schechter Ly$\alpha$ LFs from previous studies \citep{Ouchi08,Dawson07} and shown in Figure \ref{Fig_MadauPlot}. The LAE number and Ly$\alpha$ luminosity densities do not change much at $z<5.7$ but do decrease at $z>5.7$. The upper limit on number and Ly$\alpha$ luminosity densities at $z=7$ are $n_{{\rm Ly}\alpha}^{z=7}/n_{{\rm Ly}\alpha}^{z=5.7} \sim 7.7$--54\% and $\rho_{{\rm Ly}\alpha}^{z=7}/\rho_{{\rm Ly}\alpha}^{z=5.7} \sim 5.5$--39\% of those at $z=5.7$.

We have discussed differences in $n_{{\rm Ly}\alpha}$ and $\rho_{{\rm Ly}\alpha}$ between $z=7$ and later epochs based on the upper limit $z=7$ Ly$\alpha$ LF in the SXDS. This LF was derived from Ly$\alpha$ fluxes of $z=7$ LAE candidates directly converted from NB973 magnitudes without correcting for UV continuum fluxes. Even so, the $z=7$ LF shows deficit in number and Ly$\alpha$ luminosity densities from $z=5.7$. Hence, in reality, if the UV continuum fluxes are corrected, the $z=7$ LF might show more deficit. 

\subsection{Comparison with a Galaxy Evolution Model \label{KTN10}}
The deficit in densities from $z=5.7$ to $z=7$ could reflect both or either of galaxy evolution between these epochs and attenuation of Ly$\alpha$ emission by neutral hydrogen remaining at the epoch of reionization. It is possible to estimate how much the number density or Ly$\alpha$ luminosity density changes due to galaxy evolution by using empirically-calibrated semi-analytic galaxy evolution models. In Figure \ref{Fig_UpperLyaLF}, we also plot Ly$\alpha$ LFs of LAEs at $z=5.7$, 6.6 and 7 expected in the case of the IGM transmission for Ly$\alpha$ photons $T_{{\rm Ly}\alpha}^{\rm IGM}=1$ (or equivalently neutral fraction $x_{\rm HI}=0$), which we calculated by using a LAE evolution model of \citet[hereafter KTN10 model]{ktn07,ktn10}. \citet{ktn07,ktn10} constructed this model by incorporating new modeling for an escape fraction of Ly$\alpha$ photons from galaxies into a recent hierarchical clustering model of \citet{ny04}, physically considering dust extinction of Ly$\alpha$ photons and the effect of galaxy-scale outflows. The KTN10 model was empirically calibrated to fit the observed Ly$\alpha$ LF of $z=5.7$ LAEs in the SDF \citep{Shima06}. It is worth noting that, with the consistent set of model parameters, the model naturally reproduces all of the available observed data of the LAEs ({\it i.e.}, Ly$\alpha$ LF, UV LF, and equivalent width distribution) in the redshift range of $z \sim 3$--6 under the standard scenario of hierarchical galaxy formation.

The KTN10 model $z=7$ LF shows lower number density than $z=5.7$ LFs due to the galaxy evolution. The observed upper limit $z=7$ LF is consistent with the KTN10 $z=7$ LF at the bright end but shows a deficit at the fainter end. Integrating the KTN10 LF to our survey limit of $L({\rm Ly}\alpha) \gtrsim 9.2 \times 10^{42}$ erg s$^{-1}$, we obtain the number and Ly$\alpha$ luminosity densities of $n_{{\rm Ly}\alpha}^{z=7, {\rm KTN10}} \sim 7.2 \times 10^{-5}$ Mpc$^{-3}$ and $\rho_{{\rm Ly}\alpha}^{z=7, {\rm KTN10}} \sim 9.5 \times 10^{38}$ erg s$^{-1}$ Mpc$^{-3}$ as also shown in Figure \ref{Fig_MadauPlot} and Table \ref{Density}. Thus, the observed number and Ly$\alpha$ luminosity densities are factors of $n_{{\rm Ly}\alpha}^{z=7}/n_{{\rm Ly}\alpha}^{z=7, {\rm KTN10}} \sim 0.14$--0.39 and $\rho_{{\rm Ly}\alpha}^{z=7}/\rho_{{\rm Ly}\alpha}^{z=7, {\rm KTN10}} \sim 0.17$--0.36 lower than the model-predicted densities of $z=7$ LAEs in the environment where $x_{\rm HI}=0$.

This density discrepancies between the model and observed LFs might possibly reflect the attenuation of Ly$\alpha$ luminosity by neutral hydrogen. If we assume that this is the case, we can estimate the IGM transmission for Ly$\alpha$ photons $T_{{\rm Ly}\alpha}^{\rm IGM}$ at $z=7$ with $T_{{\rm Ly}\alpha}^{\rm IGM} =$ observed $L({{\rm Ly}\alpha})/$KTN10 model $L({{\rm Ly}\alpha})$. We shifted the KTN10 $z=7$ LF by attenuating its Ly$\alpha$ luminosity $L({{\rm Ly}\alpha})$ and compared it with our observed $z=7$ LF to find the possible range of $T_{{\rm Ly}\alpha}^{\rm IGM}$. In Figure \ref{Fig_UpperLyaLF}, we also show the attenuated KTN10 $z=7$ LF, which still passes through the observed $z=7$ LF data within error at the faintest end and gives the maximum possible attenuation factors of $T_{{\rm Ly}\alpha}^{\rm IGM}=0.4$. Hence, the Ly$\alpha$ luminosity is attenuated by a factor of $0.4 \lesssim T_{{\rm Ly}\alpha}^{\rm IGM} \lesssim 1$ (see also Table \ref{Density}).

\subsection{Implication for Reionization \label{XHI}}
In \textsection \ref{KTN10}, we estimated the effect of galaxy evolution on the decrease in the number density by using the KTN10 model. Even after correctiong for the galaxy evolution effect, the number density of $z=7$ LAE Ly$\alpha$ LF still shows a deficit. This could be due to Ly$\alpha$ photon absorption by neutral IGM, implying a possible evolution of the IGM neutral fraction between $z=5.7$ and $z=7$. If we can translate $T_{{\rm Ly}\alpha}^{\rm IGM}$ obtained in \textsection \ref{KTN10} to neutral fraction at $z=7$, we might be able to obtain some quantitative implications for reionization. However, this conversion process is highly complicated and uncertainty is large because we have to rely on currently proposed models that quantify Ly$\alpha$ emission line attenuation by neutral IGM \citep[{\it e.g.},][]{Santos04,Dijkstra07,McQuinn07}. As we did in \citet{Ota08} and for comparison with their result from the $z=7$ LAE survey in the SDF, we use the dynamical model with a reasonable velocity shift of the Ly$\alpha$ line by 360 km s$^{-1}$ redward of the systemic velocity \citep{Santos04}, which provides $T_{{\rm Ly}\alpha}^{\rm IGM}$ as a function of neutral fraction $x_{\rm HI}$.

\begin{figure}
\plotone{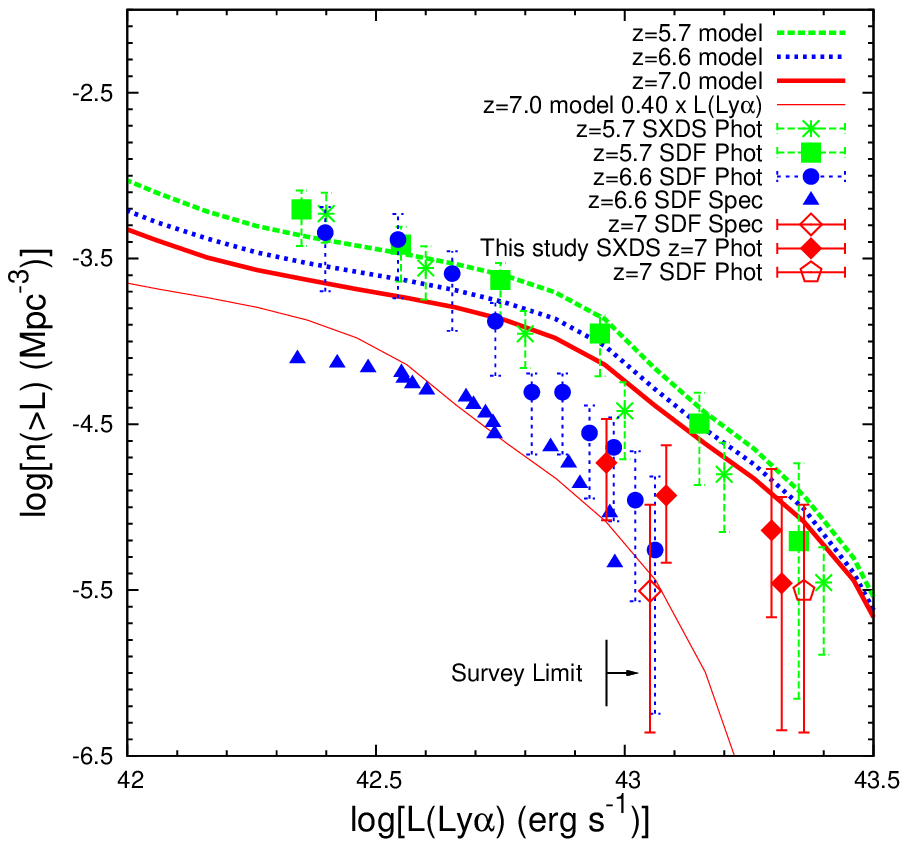}
\caption{Cumulative Ly$\alpha$ LFs of $z=5.7$ LAEs in the SDF \citep{Shima06} and SXDS \citep{Ouchi08}, $z=6.6$ LAEs \citep{kashik06} and $z=7$ LAEs in the SDF \citep{Ota08} and SXDS (this study based on the 3 candidates). The LFs based on the photmetry and spectroscopy samples are labeled as Phot and Spec. The $z=7$ SXDS LF is the upper limit LF based on the assumption that $F({\rm Ly}\alpha)=F({\rm NB973})$ in each $z=7$ LAE candidate while $L(\rm Ly\alpha)$'s of all the other LFs are corrected for the contribution from UV continuum fluxes (See esch reference). There are 4 data points for our 3 $z=7$ candidates because the last bin corresponds to our survey limit of $L_{\rm limit}({\rm Ly}\alpha)=9.2\times10^{42}$ erg s$^{-1}$ (shown by the arrow) and the faintest candidate is brighter than this limit. All the errors include cosmic variance and Poisson errors (See \textsection \ref{LyaLF}). All the data and errors are corrected for their detection completeness (See Figure \ref{MS_AVE_NB973_Completeness}). The long-dashed, short-dashed and solid curves are the intrinsic ({\it i.e.}, $T_{{\rm Ly}\alpha}^{\rm IGM}=1$) Ly$\alpha$ LFs at $z=5.7$, 6.6 and 7, respectively, predicted by the KTN10 LAE evolution model (See \textsection \ref{KTN10}). The thin solid curve is the KTN10 model $z=7$ LF with its Ly$\alpha$ luminosity attenuated by a factor of 0.4 to be consistent with the data points with its errors that result in the largest attenuation of the LF. \label{Fig_UpperLyaLF}}
\end{figure}

\begin{figure}
\plotone{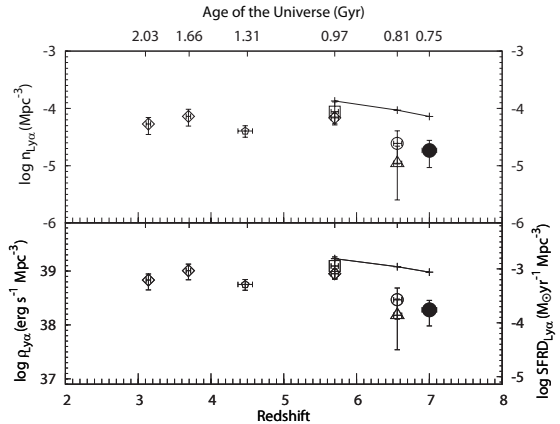}
\caption{\footnotesize The number density $n_{\rm Ly\alpha}$, Ly$\alpha$ luminosity density $\rho_{\rm Ly\alpha}$ and star formation rate density SFRD$_{\rm Ly\alpha}$ of LAEs at $z=5.7$, 6.6 and 7 derived from the current study and the latest Subaru/Suprime-Cam LAE surveys in the SDF and SXDS and those at $3 < z < 5.7$ from literature to $L_{\rm limit}({\rm Ly}\alpha)=9.2\times10^{42}$ erg s$^{-1}$. The densities at $z=7$ (large filled circles) are calculated from our 3 $z=7$ LAE candidates in the SXDS with the assumption that $F({\rm Ly}\alpha)=F({\rm NB973})$. The densities at all the other redshifts are calculated by integrating the the best-fit Ly$\alpha$ Schechter LFs derived from the $z=6.6$ SDF photometric and spectroscopic samples of \citet[open circles and triangles, respectively]{kashik06}, the $z=5.7$ SDF photometric sample of \citet[open squares]{Shima06}, the $z=3.1$, 3.7 and 5.7 SXDS samples of \citet[diamonds]{Ouchi08} and the $z \sim 4.5$ sample of \citet[pentagon]{Dawson07}. All the parameters of the Schechter LFs (normalization $\phi^{*}$, characteristic luminosity $L^{*}$ and faint end slope $\alpha$) used for the calculations of the densities shown here are the same as those used for the similar density calculations in \citet{Ota08}. Each horizontal error bar shows the redshift range of each survey. The vertical error bars include both cosmic variance and Poisson errors (See \textsection \ref{LyaLF}). The plus symbols at $z=5.7$, 6.6 and 7 with solid lines show the expected densities obtained by integrating the intrinsic Ly$\alpha$ LFs predicted by the KTN10 model. At $z>5.7$, the densities decrease and smaller than the model-predicted values, implying that the Ly$\alpha$ lines might be attenuated by the possibly increasing neutral IGM at reionization epoch.  \label{Fig_MadauPlot}}
\end{figure}

In \textsection \ref{KTN10}, we obtained $T_{{\rm Ly}\alpha}^{\rm IGM} = 0.4$--1 from the comparison of the KTN10 model LF and the upper limit on $z=7$ Ly$\alpha$ LF in Figure \ref{Fig_UpperLyaLF}. If we apply this $T_{{\rm Ly}\alpha}^{\rm IGM}$ to the Ly$\alpha$ flux attenuation model of \citet{Santos04}, the neutral fraction at $z=7$ would be $0 \lesssim x_{\rm HI}^{z=7} \lesssim 0.63$. This range of $x_{\rm HI}^{z=7}$ is consistent with $x_{\rm HI}^{z=7} \sim 0.32$--0.62 obtained by our previous NB973 survey in the SDF \citep{Ota08}, which also used the \citet{ktn07} LAE evolution model, the previous version of the KTN10 model. Since the difference in predicted number density of $z=7$ Ly$\alpha$ LF between the \citet{ktn07} model and the KTN10 model is negligibly small to our survey limit, the consistency of the result between the SXDS and SDF implies that the field-to-field variance in reionization state is not so significant between these different sky fields of comparable areas.    

Meanwhile, the Gunn-Peterson troughs \citep{GP65} found in $z \sim 6$ quasar spectra suggest that reionization ended at $z \sim 6$ with an estimated neutral fraction at $z\sim6.2$, $x_{\rm HI}^{z\sim6.2}\sim 0.01$--0.04 \citep{Fan06}. Also, spectral analyses of $z\sim6.3$ and 6.7 GRBs imply that reionization is not yet complete at $z\sim6.3$ and 6.7 with $x_{\rm HI}^{z\sim6.3}\leq 0.17$--0.6 and $x_{\rm HI}^{z\sim6.7} > 0.35$ \citep{Totani06,Greiner09}. These results, together with the $z=7$ neutral fraction we derived, imply that $x_{\rm HI}$ could be higher at the earlier epoch beyond $z\sim6$ at which reionization is believed to have ended. However, it should be again noted that our quantitative results from LAE surveys are model-dependent and thus considered one possible interpretation. All the densities, IGM transmission and neutral fraction obtained in \textsection \ref{LyaLF}--\ref{XHI} ($n_{{\rm Ly}\alpha}$, $\rho_{{\rm Ly}\alpha}$, $T_{{\rm Ly}\alpha}^{\rm IGM}$, $x_{\rm HI}^{z=7}$) are summarized in Table \ref{Density}. 

\section{Conclusion}
We conducted a deep narrowband NB973 survey of $z=7$ LAEs in the SXDS using the fully depleted CCDs newly installed on the Subaru Telescope Suprime-Cam, which doubled the detector sensitivity to $z=7$ Ly$\alpha$ emission at $\sim 1 \mu$m. Integrating 13 hours, we reached a survey limit of NB973 $=25.4$ ($5\sigma$), 0.5 mag deeper than our previous survey in the SDF. To this limit, we detected three candidate $z=7$ LAEs, all of which are detected in all the multi-epoch NB973 imaging data and thus probably real objects. 

\begin{table*}
\begin{center}
\caption{LAE Number and Ly$\alpha$ Luminosity Densities, IGM Transmission and Neutral Fraction \label{Density}}
\begin{tabular}{lcccc}\tableline\tableline
Ly$\alpha$ Luminosity Function$^a$ & $n_{{\rm Ly}\alpha}^b$ & $\rho_{{\rm Ly}\alpha}^b$ & $T^{\rm IGM}_{{\rm Ly}\alpha}$$^c$ & $x^{z=7}_{\rm HI}$$^d$ \\
              &  $(10^{-5}$ Mpc$^{-3})$   & $(10^{38}$ erg s$^{-1}$ ${\rm Mpc}^{-3})$  & &  \\ \tableline
$z=7.0$ Upper             & $1.9_{-0.9}^{+0.9}$  & $1.9_{-0.9}^{+0.9}$  &    --- &  --- \\
$z=5.7$ SDF Phot          & $8.8_{-3.6}^{+4.2}$  & $12.4_{-5.1}^{+5.9}$ &    --- &  --- \\ 
$z=5.7$ SXDS Phot         & $6.8_{-1.5}^{+1.5}$  & $9.0_{-1.9}^{+1.9}$  &    --- &  --- \\  
$z=6.6$ SDF Phot          & $2.5_{-1.4}^{+1.6}$  & $2.9_{-1.6}^{+1.9}$  &    --- &  --- \\  
$z=6.6$ SDF Spec          & $1.1_{-0.8}^{+1.1}$  & $1.5_{-1.2}^{+1.5}$  &    --- &  --- \\  
$z=7.0$ KTN10             & 7.2                  & 9.5                  &    0.40--1.00 & 0.00--0.63 \\  
\tableline
\end{tabular}
\end{center}
{$^a$These LFs correspond to the ones in Figure \ref{Fig_UpperLyaLF}.}\\
{$^b$Number and Ly$\alpha$ luminosity densities calculated to our survey limit of $L({\rm Ly}\alpha) = 9.2 \times 10^{42}$ erg s$^{-1}$ (See \textsection \ref{LyaLF} and \ref{KTN10}).}\\
{$^c$IGM transmission for Ly$\alpha$ photons at $z=7$ derived from the comparison of $z=7$ SXDS LF and $z=7$ KTN10 LF (See \textsection \ref{KTN10}).}\\
{$^d$Neutral fraction at $z=7$ obtained by appying the $T^{\rm IGM}_{{\rm Ly}\alpha}$ to \citet{Santos04} model.}
\end{table*}

With the survey depth and the sample statistics significantly improved compared with our previous NB973 survey in the SDF, we derived deeper but upper limit Ly$\alpha$ LF at $z=7$, assuming that $F({\rm NB973})=F({\rm Ly}\alpha)$. To our survey limit $L({\rm Ly}\alpha) = 9.2 \times 10^{42}$ erg s$^{-1}$ (corresponding to NB973 $=25.4$), the $z=7$ LF is similar to the $z=5.7$ LAE Ly$\alpha$ LF at the brighter end but shows a clear deficit from the $z=5.7$ at the fainter end. The fainter end of the $z=7$ LF also looks similar to the $z=6.6$ LAE Ly$\alpha$ LF. The number and Ly$\alpha$ luminosity densities at $z=7$ are $n_{{\rm Ly}\alpha}^{z=7} \sim 1.9_{-0.9}^{+0.9} \times 10^{-5}$ Mpc$^{-3}$ and $\rho_{{\rm Ly}\alpha}^{z=7} \sim 1.9_{-0.9}^{+0.9} \times 10^{38}$ erg s$^{-1}$ Mpc$^{-3}$. They are $n_{{\rm Ly}\alpha}^{z=7}/n_{{\rm Ly}\alpha}^{z=5.7} \sim 7.7$--54\% and $\rho_{{\rm Ly}\alpha}^{z=7}/\rho_{{\rm Ly}\alpha}^{z=5.7} \sim 5.5$--39\% of the densities at $z=5.7$. 

The decrease in number and Ly$\alpha$ luminosity densities from $z=5.7$ to $z=7$ could be due to either or both of galaxy evolution and suppression of Ly$\alpha$ line flux by neutral hydrogen. Our observed $z=7$ Ly$\alpha$ LF shows deficit from the predicted $z=7$ LF corrected for the possible density decrease due to galaxy evolution calculated by using a LAE evolution model (the KTN10 model). If we attribute the deficit to the attenuation of Ly$\alpha$ flux by neutral hydrogen, the IGM transmission for Ly$\alpha$ photons at $z=7$ would be $0.4 \lesssim T_{{\rm Ly}\alpha}^{\rm IGM} \lesssim 1$.    

Our current result as to $T_{{\rm Ly}\alpha}^{\rm IGM}$ strengthens the possibility that reionization might not be complete yet at $z=7$ and supports the previous results from the $z\sim6$ quasars and $z>6$ GRBs, which suggest that reionization might have completed at $z\sim6$. The Ly$\alpha$ LF, trend of decrease in $n_{{\rm Ly}\alpha}$ and $\rho_{{\rm Ly}\alpha}$ from $z=5.7$, and $T_{{\rm Ly}\alpha}^{\rm IGM}$ at $z=7$ investigated in the present deep NB973 survey in the SXDS are all consistent with those studied in our previous NB973 survey in the SDF. This implies that field-to-field variance in these properties as well as reionization status between the two different sky fields of comparable areas, the SXDS and SDF (824 and 876 arcmin$^2$, respectively, corresponding to the survey volume of $\sim 3 \times 10^5$ Mpc$^3$), is not significant. However, observing two sky fields is statisctically small and increasing the number of survey fields is still important to investigating the field variance of reionization state more accurately. 
 
The results regarding $z=7$ LAEs presented here are based on deeper survey limit and larger sample statistics than the SDF and our current conclusion is robuster. However, the sample is constructed from the photometric candidates, and the follow-up spectroscopy of them are indispensable to know how many of them are real, to estimate Ly$\alpha$ flux and luminosity precisely and thus to increase the accuracy of the study on the galaxy evolution and reionization. Even so, our current study shows that even if all the three candidates would be spectroscopically identified to be real LAEs at $z=7$, the fact that the number and Ly$\alpha$ luminosity densities decrease from $z\sim6$ to $z=7$ will not change. If some of the candidates turn out to be unreal, the densities will be even smaller. Hence, something could be really going on beyond $z\sim6$, which might possibly be reionization. This will still remain an open question for the time being, because our constraint on reionization at $z=7$ depends on one or two models: a LAE evolution model and/or a Ly$\alpha$ flux attenuation model. The improvement of the accuracy of these semi-analytical and theoretical models in the future will help answer the question.  



\acknowledgments
We are deeply grateful to the engineers of Asahi Spectra Co., Ltd.~and Hamamatsu Photonics and any people involved for developing the NB973 filter and the fully depleted CCDs for the Suprime-Cam. These technologies made the wide and deep $z=7$ LAE survey possible. We greatly appreciate the staff at the Subaru Telescope for their kind supports to make our observations successful. We express the gratitude to the SXDS team for obtaining and providing us with invaluable imaging data. We would like to thank the anonymous refree for the useful advice that helped us to improve this paper. KO acknowledges the fellowship support from the Special Postdoctoral Researchers Program at RIKEN where he conducted most of his work on this paper. MARK and TM are supported with the fellowship from the Japan Society for the Promotion of Science.




{\it Facilities:} \facility{Subaru (Suprime-Cam)}.

\end{document}